\documentclass[aps,prl,10pt,twocolumn]{revtex4-1}
\usepackage{graphicx}
\usepackage{amsmath}
\usepackage{amssymb}
\usepackage{color}
\usepackage{mathrsfs}

\begin{document}

	\title{Real-time Exciton-Manipulation of Plexcitonic Coupling in a WS$_2$-Ag nanocavity}
	
	\author{Xiaobo Han$^1$}
	
	\author{Kai Wang$^2$}
	\email{kale\_wong@hust.edu.cn}
	
	\author{Yanan Jiang$^2$, Xiangyuan Xing$^2$, Shujin Li$^2$, Fang Li$^1$,  Weiwei Liu$^2$, Bing Wang$^2$}

	\author{Peixiang Lu$^{1,2,3}$}
	\email{lupeixiang@hust.edu.cn}

	\affiliation{%
		$^1$Hubei Key Laboratory of Optical Information and Pattern Recognition, Wuhan Institute of Technology, Wuhan 430205, China\\
		$^2$Wuhan National Laboratory for Optoelectronics and School of Physics,
		Huazhong University of Science and Technology, Wuhan 430074, China\\
		$^3$CAS Center for Excellence in Ultra-intense Laser Science, Shanghai 201800, China
			}%

	\date{\today}
	
	\begin{abstract}
	We demonstrate the real-time exciton-manipulation of plexcitonic coupling in monolayer WS$_2$ coupled to a plasmonic nanocavity by immersing into a mixed solution of dichloromethane (DCM) and ethanol. By adjusting the mixture ratio, a continuous tuning of the Rabi splitting energy ranged from 178 meV (in ethanol) to 266 meV (in DCM) is achieved. The results are mainly attributed to the remarkable increase of the neutral exciton density in monolayer WS$_2$ as the concentration of DCM is increased. It offers an important stepping stone towards a further study on plexcitonic coupling in layered materials, along with potential applications in quantum information processing and nonlinear optical materials.			
	\end{abstract}                         %
	
	\maketitle
		
	Light-matter interactions are essential to many contemporary scientific disciplines \cite{1, 2, 3}. An especial interaction regime called strong coupling \cite{4,5,6}, is achieved when the rate of coherent energy exchange between an emitter and a cavity exceeds their intrinsic dissipation rates. Layered materials, in particular, two-dimensional transition metal dichalcogenides (TMDCs) possessing a huge dipole moment, attract much interest in the research of strong light-matter interaction \cite{7}. Recently, several studies demonstrate strong plasmon-exciton (plexcitonic) coupling between TMDCs and plasmonic nanostructures . Although TMDCs provide a better platform for realizing strong coupling at room temperature than quantum dots (QDs) and J-aggregates, the observed Rabi splitting (RS) energy is quite limited ($<$ 200 meV) \cite{8,10,19,12,16,11,9}. The largest RS energies are obtained to be 145 meV  and 175 meV in mono- \cite{11} and multi-layer TMDCs \cite{9} coupled to plasmonic nanostructures, and it is still smaller than that in J-aggregates system (typical value $\sim$ 400 meV) \cite{13, 14}. Therefore, it is essential to enhance plexcitonic coupling strength in TMDCs/plasmonic nanocavity.
		
	On the other hand, controlling the strong coupling strength in real-time is highly desirable for a versatile coupling system with precisely tailored optical responses \cite{50}. As recently reported, it has been demonstrated in TMDCs/plasmonic nanocavity with electric-tuning \cite{51,52}. However, the reported strong coupling strength is limited to $\sim$60 meV, while it can only be turned down to the weak coupling regime with a tuning width of $\sim$60 meV. Therefore, it is quite a challenging task to manipulate the plexcitonic coupling in real-time with an enhanced coupling strength in TMDCs/plasmonic nanostructures. In general, it is the neutral exciton that interacts with surface plasmon for strong coupling \cite{8,10,11,12,19,16,9}. Excitingly, excitons in TMDCs are proved to be reversibly converted from trions in TMDCs by chemical doping \cite{27} and solvent effects \cite{20,22}. Furthermore, the transition dipole moment in solvent-immersed TMDCs is also increased, which is conductive to enhancing strong coupling strength. Therefore, it is expected to enhance and tune the coupling strength in real-time by manipulating excitons in TMDCs under different solvents immersing. 

		\begin{figure}[hbtp]
		\centerline{
			\includegraphics[width=9cm]{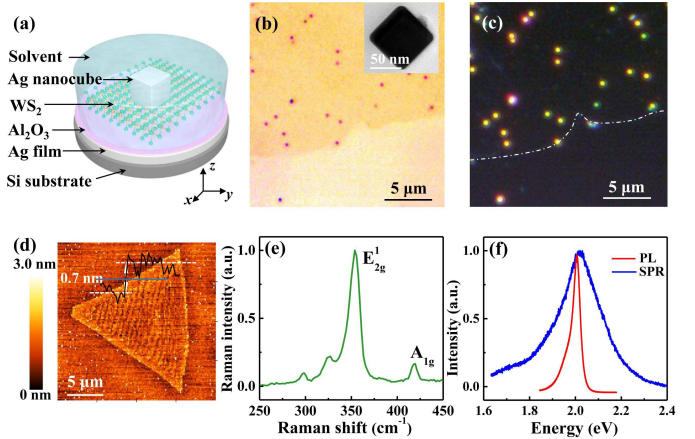}}  
		\caption{\label{fig1} (a) Scheme of a WS$_2$-Ag nanocavity immersed in a solvent. (b, c) Bright- and dark-field images of the hybrid. Inset in (b): TEM image of the Ag nanocube. (d) AFM image of a WS$_2$ flake grown on the sapphire substrate. Inset: AFM height profile indicated by the blue line. (e) Raman spectrum of monolayer WS$_2$ under 532-nm laser excitation. (f) PL spectrum of the monolayer WS$_2$ (red curve) and scattering spectrum of a plasmonic nanocavity (blue curve).
		}
	    \end{figure}

	In this letter, we demonstrate a real-time exciton-manipulation of plexcitonic coupling in monolayer WS$_2$-Ag nanocavity by immersing into the mixed solution of  dichloromethane (DCM) and ethanol. By increasing the concentration of DCM, the RS energy is tuned from 178 meV (in ethanol) to 266 meV (in DCM) continuously. It is mainly attributed that the neutral exciton density in monolayer WS$_2$ is remarkably increased as the concentration of DCM is increased. To the best of our knowledge, it shows the largest RS energy value up to 266 meV with a tuning width of 88 meV in layered TMDCs/plasmonic nanocavity. In addition, this method is very simple and convenient in technique with an ultrafast response and high reversibility. It may have potential applications in quantum information processing and manipulating chemical reaction.	
			
	The schematic diagram of the WS$_2$-Ag nanocavity is shown in Fig. 1a. The plasmonic nanocavity belongs to a highly confined nanoparticle over mirror (NPoM) geometry, which consists of a single Ag nanocube and an Ag film. A thin spacer film (Al$_2$O$_3$) and a monolayer WS$_2$ were inserted into the nanogap. The Al$_2$O$_3$ film was deposited on the steamed Ag film by atomic layer deposition (ALD), and then the monolayer WS$_2$ was transferred onto it \cite{24}. Finally, the Ag nanocubes with a polyvinylpyrrolidone (PVP) layer (nanoComposix, size of 65$\sim$95 nm) were drop-coated onto the WS$_2$ monolayers. The surface plasmon resonance (SPR) of plasmonic nanocavity can be tuned by the nanogap thickness and nanocube size. More details on the sample preparation are shown in Supporting Information (SI) \cite{41}. A custom microscopy spectroscopy (Olympus, BX 53) was used for optical measurements. The scattering signals from single hybrids were analyzed by a spectrometer (Andor, SR303i) and a CCD camera.

	Fig. 1b shows an optical image of WS$_2$-Ag nanocavity. The dark-yellow area presents a WS$_2$ flake and red dots present single Ag nanocubes. The transmission electron microscopy (TEM) image in the inset shows the Ag nanocube with an edge length of $\sim$ 75 nm. The corresponding dark-field (DF) scattering image is shown in Fig. 1c. The white dashed curve indicates the edges of WS$_2$ and the bright spots present single nanocubes. In Fig. 1d, the atomic force microscopy (AFM) image shows a triangular WS$_2$ flake. The height profile (inset) indicates that the monolayer WS$_2$ with a thickness of $\sim$ 0.7 nm is used. Raman spectrum of WS$_2$ flakes is shown in Fig. 1e, presenting two typical peaks of monolayers. The red curve shown in Fig. 1f is the photoluminescence (PL) spectrum of monolayer WS$_2$. The blue curve is the scattering spectrum of the plasmonic nanocavity, which is tuned to overlap the PL peak by adjusting the thickness of Al$_2$O$_3$ film. The quality factor, Q, of the nanocavity in air is calculated to be $\sim$10.
	
		
	\begin{figure}[hbtp]
	\centerline{
		\includegraphics[width=8.4cm]{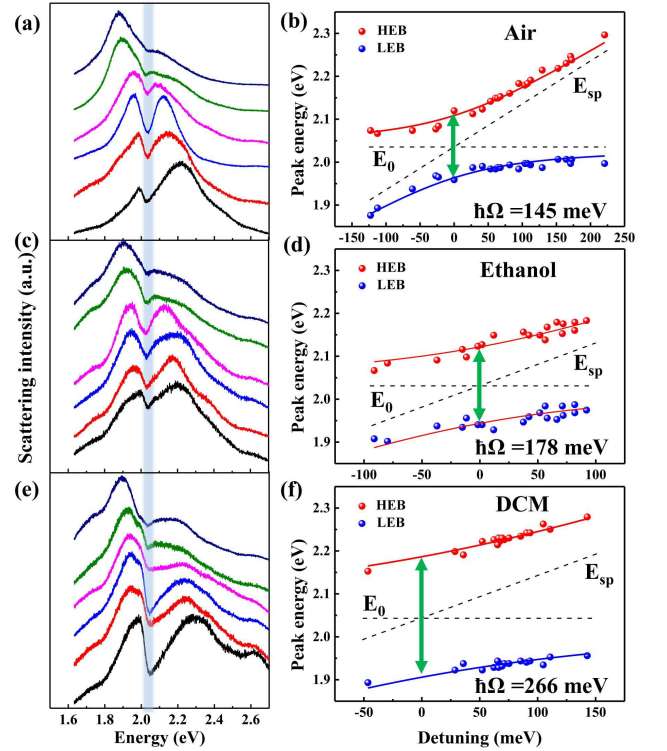}}  
	\caption{\label{fig2} (a, c, e) Scattering spectra from hybrids with different sizes of Ag nanocubes in air, ethanol and DCM, respectively. (b, d, f) Dispersions of the hybrid states in air, ethanol and DCM, respectively. It is extracted from the experimental data (red and blue dots) and calculated by using Equation (1) (red and blue lines). The black dashed lines represent the uncoupled exciton transition energy and plasmon resonance energy, respectively. 
	}
    \end{figure}
		
	Fig. 2a presents the scattering spectra of WS$_2$-Ag nanocavity with different Ag nanocube sizes in air (see Fig. S1 in SI for details \cite{41}). Extracting the energies of two peaks from scattering spectra by Lorentzian-line-fit analysis, the anti-crossing results are traced out in Fig. 2b. Specifically, energy states higher (or lower) than the exciton energy are defined as high (or low) energy branch (HEB or LEB), respectively. It can be fitted with the semi-classical coupled oscillator model (SCCOM):
            	\begin{eqnarray}
            	E_ \pm   = \frac{1}{2}\left( {E_{\rm{sp}}  + E_0  \pm \sqrt {4g^2  + \delta ^2 } } \right)
             	\end{eqnarray}	
	where $E_\pm$ are the eigen energies of the hybrid system; $\delta  = E_{{\rm{sp}}}  - E_0 	$ is the detuning energy between uncoupled surface plasmon energy ($E_{\rm sp}$) and neutral exciton energy ($E_{\rm 0} $); $E_{{\rm{sp}}}$ is obtained by $E_{{\rm{sp}}} = E_+ +  E_- - E_0 	$; $g$ is the coupling strength. The two hybrid states exhibit the plasmon-exciton dispersion characteristic with RS energy of $\hbar \Omega  = 2g =  145 $ meV, at $E_{\rm{sp}} = E_{\rm 0} $. Fig. 2c and Fig. 2e present the measured multiple scattering spectra in ethanol and in DCM, and extracted anti-crossing curves are shown in Fig. 2d and Fig. 2f, respectively. The RS energies are calculated to 178 meV and 266 meV for ethanol- and DCM-immersed hybrid systems, respectively. Obviously, the coupling strength in solvent-immersed hybrids are both enhanced in comparison to that in air. It is worth noting that the line widths of exciton and surface plasmon (in air), $\gamma\rm _0$ and $\gamma\rm_{sp}$ are extracted as $\sim$ 50 meV and $\sim$ 214 meV from Fig. 1f. In addition, the line widths of surface plasmon, $\gamma\rm_{sp}$, in ethanol and in DCM are extracted as $\sim$ 238 meV and $\sim$ 263 meV (see Fig. S2 in SI \cite{ 41}). The polariton line widths, $ {{\left( {\gamma _{\rm{sp}}  + \gamma _0 } \right)} \mathord{\left/
			{\vphantom {{\left( {\gamma _{\rm{sp}}  + \gamma _0 } \right)} 2}} \right.
			\kern-\nulldelimiterspace} 2}$, are calculated to be 132 meV (in  air), 144 meV (in ethanol) and 156 meV (in DCM). Therefore, the strong-coupling is achieved in our experiment because the polariton line widths are smaller than the corresponding RS energies \cite{25, 26}. 	
		
		\begin{figure}[hbtp]
		\centerline{
			\includegraphics[width=9cm]{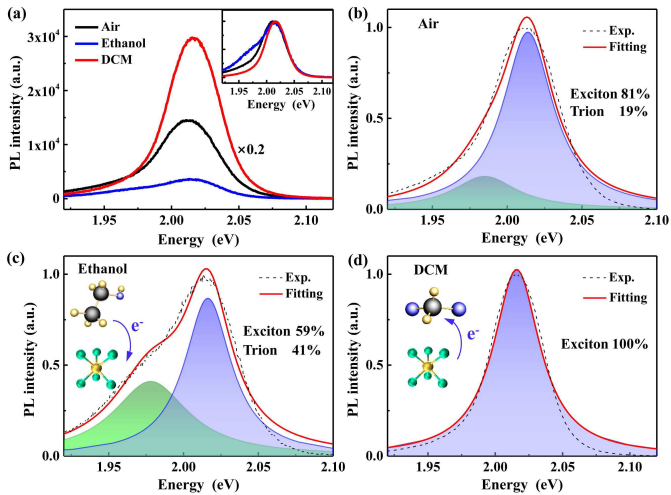}}  
		\caption{\label{fig3} (a) PL spectra of monolayer WS$_2$ on an $\rm Al_2O_3$-Ag film.  Inset: Normalized PL spectra. (b-d) PL spectra of pristine and solvent-immersed WS$_2$ (dashed black) deconvoluted into exciton ($\sim$ 615 nm; purple area) and trion ($\sim$ 626 nm; green area) peaks. The convoluted PL spectra are shown in red. The spectral weights of exciton and trion peaks are (b) 81\% and 19\%, (c) 59\% and 41\%, and (d) nearly 100\% exciton, respectively. Insets in (b and c): Artist view on electron transfer between WS$_2$ and solvents.  
		}
	\end{figure}	
	
	For analyzing the enhancement mechanism of coupling strength in different solvents, we first study the solvent effect on the PL property from monolayer WS$_2$. Fig. 3a shows PL spectra of WS$_2$ in air or immersed in different solvents under an excitation of continuous-wave laser at 473 nm. In general, it is observed that PL intensities are enhanced and quenched in DCM and ethanol solvents compared to that in air, respectively. Inset shows the normalized PL spectra at the same peak position with quite different spectral profiles. Specifically, an extra peak is observed in ethanol-immersed WS$_2$. In Fig. 3b and Fig. 3d, the spectra are analyzed by deconvoluting into two Lorentzian profiles of exciton (X at $\sim$ 615 nm) and trion (X$^-$ at $\sim$ 626 nm) peaks \cite{43}. The spectral weights of exciton and trion in monolayer WS$_2$ in air (Fig. 3b) are fitted to be 81\% and 19\%, respectively. The trion emission originates from the transition of two electrons bound to one hole, on account of abundant electrons in n-type WS$_2$ \cite{27}. In ethanol-immersed WS$_2$ (Fig. 3c), the exciton weight is decreased to 59\%, the trion weight is increased to 41\%. In contrast, DCM-immersed WS$_2$ (Fig. 3d) displays the opposite behavior. Fitting result shows that only the neutral exciton spectrum is observed ($\sim$ 100\%). The variation of exciton/trion ratio in solvent-immersed WS$_2$ is attributed to the electron transfer between solvents and WS$_2$ \cite{20}.

	For any two interacting molecules, the difference in electronegativity drives an electron transfer and determines the electron flow direction. Electronegativity ($\chi$) characterizes the ability of an atom to attract electrons in a material, which is defined as \cite{28}:	
             	\begin{eqnarray}
            	\chi _{\rm } = \frac{1}{2} \left ({\rm IP} + \rm EA \right )  	
            	\end{eqnarray}			
	where IP is the ionization potential and EA is the electron affinity. Thus, the electronegativity value of ethanol and DCM are calculated to be 2.49 and 5.05 eV, respectively \cite{30, 28, 29, 54}. The electronegativity of TMDCs can be expressed as \cite{32, 31}:
	            \begin{eqnarray}
            	\chi _{_{\rm {TMDC}}}=E_{\rm {VBM}}  + E_{\rm e } - 0.5E_{\rm g} 	
	           \end{eqnarray}	
	where $ E_{\rm {VBM}}$ is the potential of valence band maximum (VBM) against normal hydrogen electrode (NHE), $ E_{\rm e} $ is the standard electrode potential on the NHE scale, and $ E_{\rm g} $ is the bandgap.  The electronegativity value of monolayer WS$_2$ is calculated to be $\sim$ 4.83 eV \cite{33, 34}, and it is smaller than that of DCM. Therefore, monolayer WS$_2$ loses residual electrons in DCM solvent, and the spectral weight of exciton is increased up to nearly 100\%. On the contrary, monolayer WS$_2$ in ethanol attract electrons, and the exciton weight is decreased to 59\%. In experiment, the exciton weight in DCM-immersed WS$_2$ is increased by $\sim$ 1.2 times, while in ethanol-immersed one it is decreased by $\sim$ 0.7 times. Therefore, it inspires us to control the coupling strength by tuning exciton number in a mixed solution of DCM/ethanol.

	Fig. 4a exhibits the RS energy as a function of the  molar ratio. The molar ratio is defined as the mole fraction of DCM in  mixed solution of DCM/ethanol, denoted as $m$. Fig. 4b shows the exciton weight (blue dots) and relative PL intensity (red dots) averaged in several monolayer WS$_2$. As the molar ratio is increased from 0 to 0.12, the proportion of neutral exciton in the monolayer WS$_2$ is increased from 50\% to 95\%, and the measured RS energy is increased rapidly from 178 meV to 240 meV  accordingly. Therefore, the increased exciton number due to electron transfer is the dominating reason for the coupling enhancement in the molar ratio (0$\sim$0.12). As the molar ratio is further increased (0.12$\sim$0.4), the measured RS energy shows a slight decrease to 215 meV. Since the exciton weight and the PL intensity show a small variation, the decrease of RS energy may be attributed that the transition dipole moment of DCM is smaller than that of ethanol. As the molar ratio is increased from 0.4 to 1.0, the measured RS energy is increased from 215 meV to 266 meV. Fig. 4b shows that the PL intensity is increased while the exciton weight is stable, implying that the enhanced PL intensity is not due to charge transfer. According to the PL lifetime measurement, it is observed that PL lifetimes become longer as the molar ratio is increased. It is attributed to the reduction of nonradiative recombination rate due to defects passivation \cite{41,53}. Thus, it indicates that the exciton number in the molar ratio (0.4$\sim$1.0) is significantly increased.
	
		\begin{figure}[hbtp]
		\centerline{			\includegraphics[width=9cm]{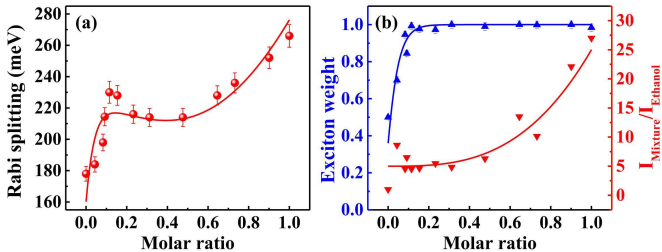}}  
		\caption{\label{fig4} (a) RS energy in a WS$_2$-Ag nanocavity as a function of the molar ratio; (b) Measured exciton weight (blue dots) and the relative PL intensity (red dot) in monolayer WS$_2$ as a function of the molar ratio. The blue and red curves indicate the numerical fitting results. 
		}
	\end{figure}

	Following is the detailed analysis of the RS energy tuning. The coupling strength is given by \cite{18, 36, 25}:
	          \begin{eqnarray}
            	g = \mu \sqrt {\frac{{4\pi \hbar Nc}}{{\lambda \varepsilon \varepsilon _0 V}}}  \propto \mu  \sqrt {{N \mathord{\left/{\vphantom {N V}} \right.
				\kern-\nulldelimiterspace} V}} 
	           \end{eqnarray}	
	where $\mu$ is the exciton transition dipole moment of monolayer WS$_2$ \cite{8}, $N$ and $\lambda $ are the number and wavelength of exciton, $ \varepsilon $ is dielectric function and $V$ is the effective mode volume of the nanocavity. Since the refractive index of nanocavity shows a low variation of 1.3\% in ethanol/DCM solvent, $V$ can be regarded as a constant for calculating coupling strength. The relative exciton transition dipole moment of TMDCs in solvents can be estimated by ${{{\mu }_{\rm{solvent}}}}/{{{\mu }_{\rm{air}}}}\;\approx \sqrt{{{\varepsilon }_{\rm {eff}}}} $ \cite{40}, where $  {\varepsilon _{\rm {eff}} }  $  is the effective dielectric constant. Thus, the relative $ \mu \left( m \right)$ as a function of molar ratio is written by:
	 \begin{eqnarray}
	 \mu \left( m \right)=m{{{{\mu }_{\rm{DCM}}}}/{{{\mu }_{\rm{air}}}}}+(1-m){{{{\mu }_{\rm{solvent}}}}/{{{\mu }_{\rm{air}}}}} 	
	\end{eqnarray}
	
	For calculating the exciton number, both of the exciton weight and PL intensity in mixed solution are considered. In Fig. 4b, exciton weight and the relative PL intensity as a function of the molar ratio of DCM/ethanol are numerically fitted, denoted as $ W $($m$) and $ I $ ($m$) (blue and red curves). Since PL intensity is proportional to the square of the exciton transition dipole moment, and it is also proportional to the exciton numbers. The exciton number ($ N $) in monolayer WS$_2$ can be calculated by: 
	
	\begin{eqnarray}
	N\left( m \right)=N_0 W\left( m \right)+{I(m)}/{{{\mu }^{2}}\left( m \right)}\;
	\end{eqnarray}
	where $ N_0 $ is the intrinsic exciton number of monolayer  WS$_2$. According to Equation (4), the RS energy can be fitted as shown in Fig. 4a (red curve), which is in good agreement with the experimental data (see SI for details). It indicates that the  tuning of coupling strength in solvents are mainly ascribed to the electron transfer and defect passivation, as well as the variation of transition dipole moment.

	In summary, we demonstrate an active exciton-manipulation of plexcitonic coupling in monolayer WS$_2$-Ag nanocavity by immersing in DCM/ethanol mixed solution. The RS energy up to 266 meV is obtained by immersing in DCM solvent, which is the largest value reported for layered TMDCs/plasmonic nanocavity, to the best of our knowledge. More importantly, the RS energy can be tuned from 145 meV (in air) to 266 meV (in DCM) by adjusting the solvent with different mixture ratio, indicating a large tuning width of 121 meV. It is mainly attributed to the remarkable increase of the neutral exciton density in monolayer WS$_2$ as the concentration of DCM is increased. This method is very simple and convenient with an ultrafast response and high reversibility. Thus, it may open a new avenue within the area of engineering strong coupling in real-time with an enhanced coupling strength in the layered TMDCs/plasmonic cavity, which has potential applications in quantum information processing and nonlinear optical materials.

    \begin{acknowledgements}
	This work was supported by NSFC (nos., 11774115, 91850113 and 11904271) and the 973 Programs under grants 2014CB92130. We thank Prof. Jun Zhou for $\rm Al_2O_3$ film fabrication. Special thanks to the Analytical and Testing Center of HUST and the Center for Nanoscale Characterization \& Devices (CNCD) of WNLO for using their facilities.
   \end{acknowledgements}

			
		
		\bibliographystyle{apsrev4-2}  
		
		\bibliography{bib4}    
	
\end{document}